\documentclass[preprint,nofootinbib,prd,aps]{revtex4}
\usepackage{graphicx,amstext,amssymb,amsmath,color}
\begin{document}

\title{ Bose-Einstein  momentum correlations at fixed
multiplicities: Lessons from an exactly solvable thermal model  for
 $pp$ collisions at the LHC}

\author{M.D. Adzhymambetov$^{1}$}
\author{S.V. Akkelin$^{1}$}
\author{Yu.M. Sinyukov$^{1}$}
\affiliation{$^1$Bogolyubov Institute for Theoretical Physics,
Metrolohichna  14b, 03143 Kyiv,  Ukraine}

\begin{abstract}

Two-particle momentum correlations of  $N$ identical bosons are
studied in the quantum canonical ensemble. We define the latter  as
a properly selected subensemble of events  associated with the grand
canonical  ensemble which is  characterized by a constant
temperature and a  harmonic-trap chemical
potential. The merits of this toy model are that it can be solved
exactly, and that it demonstrates some interesting features
revealed recently in small systems created
in $p+p$ collisions at the LHC. We find that partial coherence  can be
observed  in particle emission from completely thermal ensembles of
events if instead of inclusive measurements one studies the two-boson distribution 
functions related to the events with particle numbers selected in some fixed multiplicity bins.
The corresponding coherence effects increase with the multiplicity.

\end{abstract}


 \maketitle

\section{Introduction}

Femtoscopic study results on the  two-particle momentum correlations (see, e.g.,
Ref. \cite{Sin-1}) in $p+p$ collisions at the CERN Large Hadron
Collider (LHC) have been presented recently by the ALICE
\cite{Alice}, ATLAS \cite{Atlas},  CMS  \cite{CMS}, and LHCb \cite{LHCb} Collaborations.
It was found that the femtoscopic radii
measured by the ATLAS and CMS Collaborations decrease with
the increasing momentum of a pair. It
 can be  interpreted in the 
hydrodynamical approach  as the decrease of
``homogeneity lengths '' \cite{Sin-2} (sizes of the effective
emission region)  due to generation by the collective flow $x-p$
correlations. Also,  one found that the 
``correlation strength'' parameter $\lambda$ is essentially less than unity.
This is  at variance with the expected 
behavior for emission from thermalized  systems \cite{Sin-1}.
 
Another very interesting observation is the saturation of the multiplicity dependence of the
interferometry correlation radius parameters for very high
charged-particle multiplicity. Such an effect was observed
recently by the ATLAS \cite{Atlas} and CMS \cite{CMS} Collaborations.
Then, while there is some evidence that hydrodynamics can be successfully applied to
describe particle momentum spectra in high-multiplicity   $p+p$ collisions (for  recent review see, e.g., 
Ref. \cite{Hydro-pp-1}), it is still unclear whether the reported
results on Bose-Einstein momentum correlations can be attributed to
hydrodynamic evolution like in $A+A$ collisions.

In our opinion,
observed peculiarities of Bose-Einstein momentum correlations in
high-multiplicity  $p+p$ collisions do not indicate inapplicability
of hydrodynamics but can be partly associated with quantum coherence
effects  in small systems, when the effective system size is comparable with typical
wavelength of the thermal bosons. Recall that the effective geometrical size is associated with
the length of homogeneity in the system \cite{Sin-2}. 

Recently, a detail analysis of inclusive
spectra and Bose-Einstein correlations in small thermal quantum systems was done  for
the analytically solved model in Ref. \cite{Sin-0}.
It is shown that if one deals (even locally) with a grand canonical ensemble, a nontrivial
coherence parameter appears in inclusive two-boson spectra only in the case of
coherent condensate formation. Without the latter, no coherence-induced suppression of
the inclusive correlation function is possible because of the thermal Wick's theorem. 
    
As for nonthermal or quasithermal emission with fixed particle multiplicity,
the  traditional pair-correlation function is distorted for events with high phase-space density, 
in particular,  suppression of the Bose-Einstein correlations arises. The special algorithms for 
symmetrization of  multiboson $N$-particle states with independent particle emissions, 
and subsequent calculations of one- and two-particle spectra were developed in Refs.
\cite {Zajc, Pratt, Zhang, Urs, Ledn, Heinz}. The situation, when 
particle radiation from different source points are not independent because the wave packets of emitted bosons are overlapping,  was considered in Ref. \cite{Shap}. 
		
 Coming back to the thermal sources, in Ref. \cite{Akk-2} the coherence effects in Bose-Einstein
 correlation functions in thermal systems are studied
 in subensembles of events with fixed multiplicities. The analytical calculations were done in the 
 corresponding  canonical ensemble. It was found that the correlation functions are suppressed in a finite 
system in a large volume and low particle number density 
approximation. In the present paper, we study the two-boson momentum correlations
in small systems with high particle number densities at the moment when the system breaks up. 
Such almost sudden freeze-out  can happen due to very fast expansion (when the  homogeneity lengths are 
around $1$  fm) of the matter formed in high-multiplicity $p+p$  collisions at the LHC. 
To make the problem tractable we
utilize a model of the finite  system with smooth edges to avoid strong
boundary effects. Keeping in mind the collective expansion inherent to systems 
created in particle and nucleus collisions, 
one can associate the corresponding system's scale-parameter  with the homogeneity length.

The particle momentum spectra at a sharp freeze-out are formed according to Ref. \cite{Cooper}, which
is a reasonable approximation for
$p+p$ collisions. In order  to keep things as simple as possible, we
consider  nonrelativistic ideal gas of bosons at fixed temperature
trapped by means of a harmonic chemical potential. Such an exactly solvable
toy model of inhomogeneous and finite-sized systems is
mathematically identical to an ideal bosonic gas trapped by a
harmonic potential. Then we apply the fixed particle number
constraint to  the corresponding grand-canonical statistical
operator and discuss  the influence of such constraints on one-particle momentum
spectra and  two-boson momentum correlations.

\section{Ideal gas of bosons in a harmonic trap with fixed particle number constraint}

We begin with a brief overview of  the properties of the 
grand-canonical ensemble of  noninteracting nonrelativistic quantum-field  bosons
 at fixed temperature, $T$, trapped by a harmonic chemical potential.
For such a quantum field the Hamiltonian is given by
\begin{eqnarray}
H = \int d^{3}r \Psi^{\dag}
(\textbf{r})\left(-\frac{1}{2m}\nabla^{2}\right )\Psi(\textbf{r}),
\label{1}
\end{eqnarray}
where the operators $\Psi^{\dag} (\textbf{r})$ and $\Psi
(\textbf{r})$  are the creation and annihilation operators,
respectively. They fulfill the commutation relations
\begin{eqnarray}
[\Psi (\textbf{r}), \Psi^{\dag} (\textbf{r}') ] =
\delta^{(3)}(\textbf{r} - \textbf{r}' ), \label{2}
\end{eqnarray}
and
\begin{eqnarray}
[\Psi (\textbf{r}), \Psi (\textbf{r}')  ] =[\Psi^{\dag}
(\textbf{r}), \Psi^{\dag} (\textbf{r}')  ]= 0 . \label{3}
\end{eqnarray}
The Fourier transformed operators are defined as
\begin{eqnarray}
\Psi (\textbf{p}) = (2\pi)^{-3/2}\int d^{3}r
e^{-i\textbf{p}\textbf{r}} \Psi ( \textbf{r}), \label{4} \\
\Psi^{\dag}  (\textbf{p}) = (2\pi)^{-3/2}\int d^{3}r
e^{i\textbf{p}\textbf{r}} \Psi^{\dag} ( \textbf{r}). \label{5}
\end{eqnarray}
They satisfy the following canonical commutation relations:
\begin{eqnarray}
[\Psi ( \textbf{p}), \Psi^{\dag} (\textbf{p}') ] =
\delta^{(3)}(\textbf{p} - \textbf{p}' ),  \label{6}
\end{eqnarray}
and
\begin{eqnarray}
 [\Psi ( \textbf{p}), \Psi (\textbf{p}') ] =[\Psi^{\dag} ( \textbf{p}), \Psi^{\dag} (\textbf{p}') ]= 0.  \label{7}
\end{eqnarray}

The  grand-canonical ensemble of such a system  can be represented
by the thermal statistical operator $\rho$,
\begin{eqnarray}
\rho = \frac{1}{Z}\hat{\rho}, \label{8}
\end{eqnarray}
where $Z$ is the grand-canonical partition function,
\begin{eqnarray}
Z = Tr[\hat{\rho}], \label{9}
\end{eqnarray}
and
\begin{eqnarray}
\hat{\rho} = e^{- \beta \widehat{H} }, \label{10} \\
\widehat{H} = \int d^{3}r \Psi^{\dag}
(\textbf{r})\left(-\frac{1}{2m}\nabla^{2} - \mu (\textbf{r})\right
)\Psi(\textbf{r}), \label{11}
\end{eqnarray}
where $\beta =1/T$ is inverse temperature. The  chemical potential, $\mu (\textbf{r})$, reads
\begin{eqnarray}
\mu (\textbf{r})= -
\frac{m}{2}(\omega_{x}^{2}x^{2}+\omega_{y}^{2}y^{2}+\omega_{z}^{2}z^{2}) + \hat{\mu},
\label{12}
\end{eqnarray}
where $\hat{\mu}=\mbox{const}$. The expectation value of an  operator $O$ can be expressed as
\begin{eqnarray}
\langle  O\rangle = Tr[\rho  O]. \label{13}
\end{eqnarray}

It is well known that $\widehat{H}$ is not diagonal in momentum
(plane-wave) representation but can be diagonalized in the
oscillator representation. Decomposing $\Psi(\textbf{r})$ and
$\Psi^{\dag}(\textbf{r})$ in terms of the harmonic oscillator
eigenfunctions we get
\begin{eqnarray}
\Psi(\textbf{r})=
\sum_{n,k,l=0}^{\infty}\alpha(n,k,l)\phi_{n}(x)\phi_{k}(y)\phi_{l}(z),
\label{14}
\end{eqnarray}
where the creation, $\alpha^{\dag}(n,k,l)$, and annihilation,
$\alpha(n,k,l)$, operators satisfy the commutation relations
\begin{eqnarray}
[\alpha(n,k,l),\alpha^{\dag}(n',k',l')] = \delta_{nn'}
\delta_{kk'}\delta_{ll'} , \label{15}
\end{eqnarray}
and
\begin{eqnarray}
 [\alpha(n,k,l),\alpha(n',k',l')] =[\alpha^{\dag}(n,k,l),\alpha^{\dag}(n',k',l')]= 0 . \label{16}
\end{eqnarray}
Functions   $\phi_{n}(x)$, $\phi_{k}(y)$, $\phi_{l}(z)$ are the
harmonic oscillator  eigenfunctions satisfying corresponding
equations, e.g.,
\begin{eqnarray}
\left( \frac{d^{2}}{dx^{2}}-m \omega_{x}^{2}x^{2}+ 2m
\epsilon_{n}\right ) \phi_{n}(x) =0 .\label{17}
\end{eqnarray}
The normalized solution of Eq. (\ref{17}) reads
\begin{eqnarray}
 \phi_{n}(x) =(2^{n}n!\pi^{1/2}b_{x})^{-1/2}H_{n}
 \left(\frac{x}{b_{x}}\right)\exp\left( -\frac{1}{2}\left( \frac{x}{b_{x}}\right)^{2}\right),\label{18}
\end{eqnarray}
where $H_{n}(x/b_{x})$ is the Hermite polynomial, and
\begin{eqnarray}
\epsilon_{n}=\omega_{x}\left(n+\frac{1}{2}\right), \label{19} \\
b_{x}=(m\omega_{x})^{-1/2}. \label{20}
\end{eqnarray}
Eigenfunctions (\ref{18}) are complete,
\begin{eqnarray}
\sum_{n=0}^{\infty}\phi_{n}(x)\phi_{n}^{*}(x') = \delta(x-x'),
\label{21}
\end{eqnarray}
and orthonormal
\begin{eqnarray}
 \int_{-
\infty}^{\infty}\phi_{n}(x)\phi_{n'}^{*}(x)dx= \delta_{nn'}.
\label{22}
\end{eqnarray}
Then, from Eq. (\ref{14}) it immediately follows that
\begin{eqnarray}
\alpha(n,k,l)= \int_{- \infty}^{\infty}dx dy dz
\phi_{n}^{*}(x)\phi_{k}^{*}(y)\phi_{l}^{*}(z)\Psi(\textbf{r}).
\label{23}
\end{eqnarray}
In such a basis the $\widehat{H}$ reads
\begin{eqnarray}
\widehat{H} =
\sum_{n,k,l=0}^{\infty}(\epsilon_{n}+\epsilon_{k}+\epsilon_{l}-\widehat{\mu})\alpha^{\dag}(n,k,l)
\alpha(n,k,l). \label{24}
\end{eqnarray}

Equation (\ref{24}) allows one to calculate expectation values
(\ref{13}) for products of  $\alpha^{\dag}$ and $\alpha$ operators.
It can be done in various ways. It is more appropriate here to use
the method which was used to prove  the Wick's theorem for the grand-canonical ensemble  
(see, e.g., Ref. \cite{Wick})  as the extension of
it can be used for the case of the canonical ensemble. First, using
the eigenstates\footnote{For notational simplicity, here and below
we write $\textbf{j}$ instead of $(n,k,l)$.}
\begin{eqnarray}
|\textbf{j}_{1},...,\textbf{j}_{N}\rangle =
\frac{1}{\sqrt{N!}}\alpha^{\dag}(\textbf{j}_{1})...
\alpha^{\dag}(\textbf{j}_{N})|0\rangle  \label{25}
\end{eqnarray}
of the particle number operator
$\sum_{\textbf{j}}\alpha^{\dag}(\textbf{j})\alpha(\textbf{j})$, and
the identity
\begin{eqnarray}
\sum_{N=0}^{\infty}\sum_{\textbf{j}_{1}=\textbf{0}}^{\infty}...\sum_{\textbf{j}_{N}=\textbf{0}}^{{\infty}}|\textbf{j}_{1},...,\textbf{j}_{N}\rangle
\langle \textbf{j}_{1},...,\textbf{j}_{N}|= 1, \label{26}
\end{eqnarray}
which express the completeness and normalization of this basis, one
can insert  Eq. (\ref{24}) into Eq. (\ref{10}) and write
$\hat{\rho }$  in  the harmonic oscillator basis,
\begin{eqnarray}
\hat{\rho }=
\sum_{N}\sum_{\textbf{j}_{1}}...\sum_{\textbf{j}_{N}} e^{-\beta
(\epsilon_{\textbf{j}_1}-\hat{\mu})}...e^{-\beta (\epsilon_{\textbf{j}_N}-\hat{\mu})}
|\textbf{j}_{1},...,\textbf{j}_{N} \rangle \langle
\textbf{j}_{1},...,\textbf{j}_{N}| . \label{27}
\end{eqnarray}
We denote here
\begin{eqnarray}
\epsilon_{\textbf{j}}=\epsilon_{n,k,l}=\epsilon_{n}+\epsilon_{k} +
\epsilon_{l} = \omega_{x}\left(n+\frac{1}{2}\right)+
\omega_{y}\left(k+\frac{1}{2}\right)+\omega_{z}\left(l+\frac{1}{2}\right).
\label{28}
\end{eqnarray}
Then, using an elementary operator algebra and Eq. (\ref{27}) one
can see that
\begin{eqnarray}
\alpha(\textbf{j}) \hat{\rho} = \hat{\rho}
\alpha(\textbf{j}) e^{-\beta (\epsilon_{\textbf{j}}-\hat{\mu})} . \label{29}
\end{eqnarray}
Using trace invariance under the cyclic permutation of an operator,
we get
\begin{eqnarray}
Tr[\hat{\rho
}\alpha^{\dag}(\textbf{j}_{1})\alpha(\textbf{j}_{2})] = \nonumber \\
e^{-\beta ( \epsilon_{\textbf{j}_{2}}-\hat{\mu})} Tr[\hat{\rho}
\alpha(\textbf{j}_{2}) \alpha^{\dag}(\textbf{j}_{1})]= e^{-\beta (
\epsilon_{\textbf{j}_{2}}-\hat{\mu})}(Tr[\hat{\rho}
\alpha^{\dag}(\textbf{j}_{1})\alpha(\textbf{j}_{2})]+
\delta_{\textbf{j}_{1}\textbf{j}_{2}} Tr[\hat{\rho }]).
\label{30}
\end{eqnarray}
The Kronecker delta in the above equation,
$\delta_{\textbf{j}_{1}\textbf{j}_{2}}$,  is
\begin{eqnarray}
\delta_{\textbf{j}_{1}\textbf{j}_{2}}=\delta_{n_1n_2}\delta_{k_1k_2}\delta_{l_1l_2}
. \label{31}
\end{eqnarray}
From  Eq. (\ref{30}) we have
\begin{eqnarray}
\langle \alpha^{\dag}(\textbf{j}_{1})\alpha(\textbf{j}_{2})\rangle
=\frac{1}{Tr[\hat{\rho}]}Tr[\hat{\rho}\alpha^{\dag}(\textbf{j}_{1})\alpha(\textbf{j}_{2})]=
\frac{\delta_{\textbf{j}_{1}\textbf{j}_{2}}}{e^{\beta(
\epsilon_{\textbf{j}_{2}}-\hat{\mu})}-1 }, \label{32}
\end{eqnarray}
which  is a familiar Bose-Einstein distribution. It follows then that 
\begin{eqnarray}
\langle N\rangle =  \sum_{\textbf{j}} \langle \alpha^{\dag}(\textbf{j})\alpha(\textbf{j})\rangle .
 \label{32.1}
\end{eqnarray}
In a similar way,
one can get
\begin{eqnarray}
Tr[\hat{\rho
}\alpha^{\dag}(\textbf{j}_{1})\alpha^{\dag}(\textbf{j}_{2})
\alpha(\textbf{j}_{3})\alpha(\textbf{j}_{4})] =  \nonumber \\
e^{-\beta (
\epsilon_{\textbf{j}_{4}}-\hat{\mu})}(\delta_{\textbf{j}_1\textbf{j}_4}
Tr[\hat{\rho }\alpha^{\dag}(\textbf{j}_{2})
\alpha(\textbf{j}_{3})]+
\delta_{\textbf{j}_2\textbf{j}_4}Tr[\hat{\rho
}\alpha^{\dag}(\textbf{j}_{1}) \alpha(\textbf{j}_{3})] +
Tr[\hat{\rho
}\alpha^{\dag}(\textbf{j}_{1})\alpha^{\dag}(\textbf{j}_{2})
\alpha(\textbf{j}_{3})\alpha(\textbf{j}_{4})]), \label{33}
\end{eqnarray}
Then,  taking into account Eq. (\ref{32}) we have
\begin{eqnarray}
\langle \alpha^{\dag}(\textbf{j}_{1})\alpha^{\dag}(\textbf{j}_{2})
\alpha(\textbf{j}_{3})\alpha(\textbf{j}_{4})\rangle = \nonumber
\\ \langle \alpha^{\dag}(\textbf{j}_{2})
\alpha(\textbf{j}_{3})\rangle \langle \alpha^{\dag}(\textbf{j}_{1})
\alpha(\textbf{j}_{4})\rangle + \langle
\alpha^{\dag}(\textbf{j}_{1}) \alpha(\textbf{j}_{3})\rangle \langle
\alpha^{\dag}(\textbf{j}_{2}) \alpha(\textbf{j}_{4})\rangle ,
\label{34}
\end{eqnarray}
which is nothing but the particular case of the  thermal Wick's
theorem. Then, utilizing Eq. (\ref{14})  and Eqs. (\ref{32}) and
(\ref{34}) one can calculate expectation values of
$\Psi$ and $\Psi^{\dag}$ operators.

Now, let us apply the fixed particle number constraint to the
grand-canonical statistical operator (\ref{8}) to define
canonical statistical operator $\rho_{N}$.  For this aim, one can
utilize the projection operator ${\cal P}_{N}$,
\begin{eqnarray}
{\cal P}_{N} = \frac{1}{N!}\int d^{3}r_{1}...
d^{3}r_{N}\Psi^{\dag}(\textbf{r}_{1})...
\Psi^{\dag}(\textbf{r}_{N})|0\rangle \langle
0|\Psi(\textbf{r}_{1})... \Psi(\textbf{r}_{N}) , \label{35}
\end{eqnarray}
which  automatically invokes the corresponding constraint. Using
Eqs. (\ref{14}), (\ref{22}) and (\ref{25}) one can see that
\begin{eqnarray}
{\cal P}_{N} = \sum_{\textbf{j}_{1}}...
\sum_{\textbf{j}_{N}}|\textbf{j}_{1},...,\textbf{j}_{N}\rangle
\langle \textbf{j}_{1},...,\textbf{j}_{N} | .  \label{36}
\end{eqnarray}
 It is worth
noting that such a projection is accompanied by the proper
normalization in order to insure the probability interpretation of
the ensemble obtained in result of this projection. Then, using (\ref{36}) we assert that the canonical statistical
operator is
\begin{eqnarray}
\rho_{N} = \frac{1}{Z_{N}} \hat{\rho}_{N}, \label{37}
\end{eqnarray}
where
\begin{eqnarray}
\hat{\rho}_{N} = {\cal P}_{N} \hat{\rho }{\cal P}_{N}
=\sum_{\textbf{j}_{1}}... \sum_{\textbf{j}_{N}}e^{-\beta
(\epsilon_{\textbf{j}_1}-\hat{\mu})} ... e^{-\beta(
\epsilon_{\textbf{j}_N}-\hat{\mu})}|\textbf{j}_{1},...,\textbf{j}_{N}\rangle
\langle \textbf{j}_{1},...,\textbf{j}_{N} | , \label{38}
\end{eqnarray}
and  $Z_{N} $ is the corresponding  canonical partition function,
\begin{eqnarray}
Z_{N} =Tr[\hat{\rho}_{N}]. \label{39}
\end{eqnarray}
It follows from Eq. (\ref{38}) that 
\begin{eqnarray}
Z=\sum_{N=0}^{\infty} Z_{N}. \label{39.0}
\end{eqnarray}
The vacuum state, $N=0$, yields  $Z_{0} = \langle  0|0 \rangle =1$. Let us
denote $\hat{\rho}_{N}$ associated with $\hat{\mu}=0$ as $\hat{\rho}_{N}^{0}$.
Then one can readily see that $\widehat{\rho}_{N}=e^{\beta \widehat{\mu} N}\hat{\rho}_{N}^{0}$ and
\begin{eqnarray}
Z_{N} =e^{\beta \hat{\mu} N}Z_{N}^{0}. \label{39.00}
\end{eqnarray}
 Therefore [see Eq. (\ref{37})] $e^{\beta \hat{\mu} N}$ 
is factored out and $\rho_{N}$
does not depend on $\hat{\mu}$: 
\begin{eqnarray}
\rho_{N} = \frac{1}{Z_{N}^{0}} \hat{\rho}_{N}^{0} .  \label{39.1}
\end{eqnarray}
The expectation value of an operator $O$ is defined as
\begin{eqnarray}
\langle  O \rangle_{N} = Tr[\rho_{N}  O] . \label{40}
\end{eqnarray}
It follows from Eqs. (\ref{38}) and (\ref{40}) that
\begin{eqnarray}
\langle  O \rangle=\sum_{N=0}^{\infty}\frac{Z_{N}}{Z}\langle  O \rangle_{N} . \label{40.1}
\end{eqnarray}

To evaluate  the expectation values of operators
$\alpha^{\dag}(\textbf{j}_{1})\alpha(\textbf{j}_{2})$ and
$\alpha^{\dag}(\textbf{j}_{1})\alpha^{\dag}(\textbf{j}_{2})\alpha(\textbf{j}_{3})\alpha(\textbf{j}_{4})$
with the canonical statistical operator $\rho_{N}$,  one can  adopt   the
procedure which was used above to calculate expectation values with
the grand-canonical statistical operator $\rho$. It can be done  in
a similar way as it was done, e.g., in Ref. \cite{Akk-2}. For the
reader's convenience, below we adjust the derivation from Ref.
\cite{Akk-2} for our model. A starting point is the relation
\begin{eqnarray}
\alpha(\textbf{j}) \hat{\rho}_{N}^{0} = \hat{\rho}_{N-1}^{0}
\alpha(\textbf{j}) e^{-\beta \epsilon_{\textbf{j}}} \label{41}
\end{eqnarray}
which follows from Eq. (\ref{38}) and commutation relations
(\ref{15}) and (\ref{16}). Then one can exploit invariance under
cyclic permutation and get the iteration equation
\begin{eqnarray}
\langle
\alpha^{\dag}(\textbf{j}_{1})\alpha(\textbf{j}_{2})\rangle_{N} =
e^{-\beta
\epsilon_{\textbf{j}_{2}}}\delta_{\textbf{j}_1\textbf{j}_2}\frac{Z_{N-1}^{0}}{Z_{N}^{0}}+
e^{-\beta  \epsilon_{\textbf{j}_{2}}}\frac{Z_{N-1}^{0}}{Z_{N}^{0}}\langle
\alpha^{\dag}(\textbf{j}_{1})\alpha(\textbf{j}_{2})\rangle_{N-1}.
\label{42}
\end{eqnarray}
With the starting value $\langle
\alpha^{\dag}(\textbf{j}_{1})\alpha(\textbf{j}_{2})\rangle_{0} =0$
one can get from the above equation that
\begin{eqnarray}
\langle
\alpha^{\dag}(\textbf{j}_{1})\alpha(\textbf{j}_{2})\rangle_{N} =
\delta_{\textbf{j}_1\textbf{j}_2} \sum_{s=1}^{N} e^{-s\beta 
\epsilon_{\textbf{j}_{2}}}\frac{Z_{N-s}^{0}}{Z_{N}^{0}}. 
\label{43}
\end{eqnarray}
It follows from the definition of $\rho_{N}$ [see Eqs. (\ref{37})
and (\ref{38})] that
\begin{eqnarray}
\sum_{\textbf{j}}\langle
\alpha^{\dag}(\textbf{j})\alpha(\textbf{j})\rangle_{N}  = N.
\label{43.1}
\end{eqnarray}
Utilizing relation (\ref{41}) we  have
\begin{eqnarray}
\langle
\alpha^{\dag}(\textbf{j}_{1})\alpha^{\dag}(\textbf{j}_{2})\alpha(\textbf{j}_{3})\alpha(\textbf{j}_{4})\rangle_{N}
= e^{-\beta \epsilon_{\textbf{j}_{4}}} \frac{Z_{N-1}^{0}}{Z_{N}^{0}}\langle
\alpha(\textbf{j}_{4})\alpha^{\dag}(\textbf{j}_{1})\alpha^{\dag}(\textbf{j}_{2})\alpha(\textbf{j}_{3})\rangle_{N-1}.
 \label{44.0}
\end{eqnarray}
Then the same procedure leads to
\begin{eqnarray}
\langle
\alpha^{\dag}(\textbf{j}_{1})\alpha^{\dag}(\textbf{j}_{2})\alpha(\textbf{j}_{3})\alpha(\textbf{j}_{4})\rangle_{N}
=  e^{-\beta \epsilon_{\textbf{j}_{4}}}
\frac{Z_{N-1}^{0}}{Z_{N}^{0}} \times \nonumber \\ \left ( \langle
\alpha^{\dag}(\textbf{j}_{1})\alpha^{\dag}(\textbf{j}_{2})\alpha(\textbf{j}_{3})\alpha(\textbf{j}_{4})\rangle_{N-1}
+  \delta_{\textbf{j}_1\textbf{j}_4} \langle
\alpha^{\dag}(\textbf{j}_{2})\alpha(\textbf{j}_{3})\rangle_{N-1} +
\delta_{\textbf{j}_2\textbf{j}_4} \langle
\alpha^{\dag}(\textbf{j}_{1})\alpha(\textbf{j}_{3})\rangle_{N-1} \right).
 \label{44}
\end{eqnarray}
One can show by induction that  Eq. (\ref{44}) can be written  as
\begin{eqnarray}
\langle
\alpha^{\dag}(\textbf{j}_{1})\alpha^{\dag}(\textbf{j}_{2})\alpha(\textbf{j}_{3})\alpha(\textbf{j}_{4})\rangle_{N}
= \nonumber \\ \delta_{\textbf{j}_1\textbf{j}_4} \sum_{s=1}^{N}
e^{-s \beta \epsilon_{\textbf{j}_{4}}}\frac{Z_{N-s}^{0}}{Z_{N}^{0}} \langle
\alpha^{\dag}(\textbf{j}_{2})\alpha(\textbf{j}_{3})\rangle_{N-s} +
\delta_{\textbf{j}_2\textbf{j}_4} \sum_{s=1}^{N}e^{-s \beta 
\epsilon_{\textbf{j}_{4}}}\frac{Z_{N-s}^{0}}{Z_{N}^{0}} \langle
\alpha^{\dag}(\textbf{j}_{1})\alpha(\textbf{j}_{3})\rangle_{N-s}.
 \label{46}
\end{eqnarray}
Then, taking into account that $\langle
\alpha^{\dag}(\textbf{j}_{1})\alpha(\textbf{j}_{2})\rangle_{0} =0$
and Eq. (\ref{43}),  we get
\begin{eqnarray}
\langle
\alpha^{\dag}(\textbf{j}_{1})\alpha^{\dag}(\textbf{j}_{2})\alpha(\textbf{j}_{3})\alpha(\textbf{j}_{4})\rangle_{N}
= \nonumber \\ 
(\delta_{\textbf{j}_1\textbf{j}_4}\delta_{\textbf{j}_2\textbf{j}_3}
+
\delta_{\textbf{j}_1\textbf{j}_3}\delta_{\textbf{j}_2\textbf{j}_4})\sum_{s=1}^{N-1}
\sum_{s'=1}^{N-s} e^{-s\beta \epsilon_{\textbf{j}_{4}}}e^{-s' \beta 
\epsilon_{\textbf{j}_{3}}}\frac{Z_{N-s-s'}^{0}}{Z_{N}^{0}}.
\label{47}
\end{eqnarray}
The above expressions explicitly demonstrate deviations from the
Wick's theorem in the canonical ensemble for a system of
noninteracting bosons.

Canonical partition functions in Eqs. (\ref{43})
and (\ref{47}) can be calculated 
by means of the recursive formula of the canonical partition function
for a system of $N$ noninteracting bosons  as given in Ref. \cite{Recurr-1} (an elementary derivation of it can be
seen  in Ref. \cite{Akk-2}):
\begin{eqnarray}
n Z_{n}^{0}= \sum_{s=1}^{n} \sum_{\textbf{j}} e^{-s\beta \epsilon_{\textbf{j}}}Z_{n-s}^{0},
\label{48}
\end{eqnarray}
where $Z_{0}^{0}=\langle 0 | 0 \rangle =1$ and $n=1,...,N$.

As a final comment we would like to point out that there is an essential  difference between
states defined by the  grand-canonical statistical operator, $\rho$ 
[see Eqs. (\ref{8}), (\ref{9}), and (\ref{27})] and the canonical statistical
operator, $\rho_{N}$  [see  Eqs. (\ref{37}), (\ref{38}), and (\ref{39})].
While the former is a mixture of all $N$-particle states including
vacuum state with $N=0$, the latter is a mixture  of states with  $N$
fixed to some value. In a sense, the quantum canonical state, $\rho_{N}$, can be interpreted as 
a state  which is not completely chaotic but has some quantum coherent
properties. In what follows we demonstrate that such a coherence is enhanced in
the case of the Bose-Einstein 
condensation, when the number of particles   in the ground state, $N_{0}$, 
is of the order 
of the total number of particles, $N$,\footnote{This is the  definition of the
Bose-Einstein condensation; see, e.g., 
Ref. \cite{BEC-2}.} and  discuss possible relations of our
results to two-boson momentum correlations measured in $p+p$ collisions at
the LHC.

\section{Particle momentum spectra and  correlations  at fixed multiplicities}

In this section we relate the model with physical
observables in relativistic particle and nucleus collisions.   To keep things as
simple as possible, below we assume that
$\omega_{x}=\omega_{y}=\omega_{z}\equiv \omega$. Note that the  mean
particle number, $\langle N \rangle$, defined by the  grand
canonical ensemble, as well as the  particle number, $N$, in the
canonical ensemble are the same for  $\Psi$ particles   and
$\alpha$ quasiparticles  because unitary transformation (\ref{14})
does not mix creation and annihilation operators.

First,  let us estimate spatial  size of the system at fixed
multiplicities. It is defined as $\sqrt{\frac{1}{3}\langle \textbf{r}^{2} \rangle_{N}}$, where
\begin{eqnarray}
\frac{1}{3}\langle \textbf{r}^{2} \rangle_{N}=\langle x^{2} \rangle_{N} =\frac{\int dxdydz
x^{2}\langle \Psi^{\dag}(\textbf{r}) \Psi
(\textbf{r})\rangle_{N} }{\int dxdydz \langle
\Psi^{\dag}(\textbf{r}) \Psi (\textbf{r})\rangle_{N} }, \label{49}
\end{eqnarray}
 $\langle \Psi^{\dag}(\textbf{r}) \Psi (\textbf{r})\rangle_{N}$ is the mean particle number
 density in the canonical ensemble, and $\int dxdydz \langle
\Psi^{\dag}(\textbf{r}) \Psi (\textbf{r})\rangle_{N} = N$.
 From Eqs. (\ref{14}) and (\ref{43}) we get
\begin{eqnarray}
\langle \Psi^{\dag}(\textbf{r}_{1}) \Psi (\textbf{r}_{2})\rangle_{N}
= \nonumber \\ \sum_{s=1}^{N}\frac{Z_{N-s}^{0}}{Z_{N}^{0}}
\sum_{n=0}^{\infty}\sum_{k=0}^{\infty}\sum_{l=0}^{\infty}\phi_{n}^{*}(x_{1})\phi_{k}^{*}(y_{1})\phi_{l}^{*}(z_{1})
\phi_{n}(x_{2})\phi_{k}(y_{2})\phi_{l}(z_{2})e^{-\frac{3}{2}s\beta
\omega}e^{-s \beta \omega (n+k+l)}, \label{50}
\end{eqnarray}
where the eigenfunctions  are defined by Eq. (\ref{18}), $b_{x}=b_{y}=b_{z} \equiv   b$ and 
\begin{eqnarray}
b= (m\omega)^{-1/2} , \label{50.1}
\end{eqnarray}
see Eq. (\ref{20}).
Then, utilizing integral representation of the Hermite function (see
e.g. Ref. \cite{math}),
\begin{eqnarray}
H_{n}\left( \frac{x}{b}\right )= \left ( \frac{b}{i}\right
)^{n}\frac{b}{2\sqrt{\pi}}e^{\frac{x^{2}}{b^{2}}}
\int_{-\infty}^{+\infty}v^{n}e^{-\frac{1}{4}b^{2}v^{2}+ixv}dv ,
\label{51}
\end{eqnarray}
one can perform summations over $n,k,l$ in Eq. (\ref{50}). A lengthy
but straightforward calculation  results in
\begin{eqnarray}
\langle \Psi^{\dag}(\textbf{r}_{1}) \Psi (\textbf{r}_{2})\rangle_{N}
=
\nonumber \\
\sum_{s=1}^{N}\frac{1}{(2\pi)^{3/2}}\frac{1}{b^{3}}\frac{Z_{N-s}^{0}}{Z_{N}^{0}}\left
(\sinh(\beta \omega s)\right )^{-3/2}\exp\left ( -
\frac{\textbf{r}_{1}^{2}+\textbf{r}_{2}^{2}}{2b^{2}\tanh(\beta
\omega s) } \right )\exp\left(
\frac{\textbf{r}_{1}\textbf{r}_{2}}{b^{2}\sinh(\beta \omega
s)}\right ) . \label{52}
\end{eqnarray}
 Utilizing  identity $(\tanh A)^{-1}- (\sinh
A)^{-1}=\tanh ( A/2 )$, we have from Eq. (\ref{52}) that
mean particle number density in the canonical ensemble reads
\begin{eqnarray}
\langle \Psi^{\dag}(\textbf{r}) \Psi (\textbf{r})\rangle_{N} =
\sum_{s=1}^{N}\frac{1}{(2\pi)^{3/2}}\frac{1}{b^{3}}\frac{Z_{N-s}^{0}}{Z_{N}^{0}}\left
(\sinh(\beta \omega s)\right )^{-3/2}\exp\left ( -
\frac{\tanh(\frac{1}{2}\beta \omega s) }{b^{2}} \textbf{r}^{2}\right
) . \label{54}
\end{eqnarray}
Substituting the above expression  in Eq. (\ref{49}) we readily find
\begin{eqnarray}
\langle x^{2} \rangle_{N} =
\frac{1}{2}b^{2}\frac{\sum_{s=1}^{N}\frac{Z_{N-s}^{0}}{Z_{N}^{0}}\left (
\sinh(\beta \omega s)\right )^{-3/2}\left ( \tanh(\frac{1}{2}\beta
\omega s) \right )^{-5/2}}{\sum_{s=1}^{N}\frac{Z_{N-s}^{0}}{Z_{N}^{0}}\left
( \sinh(\beta \omega s)\right )^{-3/2}\left ( \tanh(\frac{1}{2}\beta
\omega s) \right )^{-3/2}}. \label{55}
\end{eqnarray}

To relate parameters of the model with physically meaningful  
parameters in relativistic particle and nucleus collisions, it is convenient to 
introduce parameter  $R$ such as   
\begin{eqnarray}
\omega= \frac{1}{R \sqrt{\beta m}},
 \label{56}
\end{eqnarray}
then  $\frac{m \omega^{2}}{2}=\frac{1}{2\beta R^{2}}$; see Eq. (\ref{12}).
In what follows we treat $R$ as free parameter instead of $\omega$. As we will see below,
$R$ can be approximately associated with the spatial size of the system, $\sqrt{\langle x^{2} \rangle_{N}}$. 

Then 
\begin{eqnarray}
\beta \omega  = \frac{1}{R} \sqrt{\frac{\beta}{m}}=\frac{\Lambda_{T}}{R},
 \label{57}
\end{eqnarray}
and 
\begin{eqnarray}
b = \frac{1}{\sqrt{m\omega}}=\sqrt{\Lambda_{T} R},
 \label{57.1}
\end{eqnarray}
where $\Lambda_{T}$ is the thermal wavelength,  which we defined as
\begin{eqnarray}
\Lambda_{T} = \frac{1}{\sqrt{mT}}. 
 \label{58}
\end{eqnarray}

We now turn to  the two-particle momentum correlation  functions.
Two-particle momentum correlation function is defined as ratio of
two-particle momentum spectrum to one-particle ones and can be
written in canonical ensemble  at fixed multiplicities as
\begin{eqnarray}
C_{N}(\textbf{k},\textbf{q})=G_{N} \frac{\langle
\Psi^{\dag}(\textbf{p}_{1})\Psi^{\dag}(\textbf{p}_{2})\Psi(\textbf{p}_{1})\Psi(\textbf{p}_{2})\rangle_{N}}{\langle
\Psi^{\dag}(\textbf{p}_{1})\Psi(\textbf{p}_{1})\rangle_{N}\langle
\Psi^{\dag}(\textbf{p}_{2})\Psi(\textbf{p}_{2})\rangle_{N}}.
\label{59}
\end{eqnarray}
 Here ${\bf k}=({\bf
p}_{1}+{\bf p}_{2})/2$, ${\bf q}={\bf p}_{2}-{\bf p}_{1}$, and
$G_{N}$ is the normalization constant. The latter  is needed  to
normalize the theoretical correlation function in accordance with
normalization that is applied by experimentalists:
$C^{exp}(\textbf{k},\textbf{q}) \rightarrow
 1$ for $|\textbf{q}| \rightarrow \infty$.

Expressions in the denominator of Eq. (\ref{59})  can be written
immediately  using  Fourier transform of $\langle
\Psi^{\dag}(\textbf{r}_{1}) \Psi (\textbf{r}_{2})\rangle_{N}$; see
Eq. (\ref{52}). We thus have
\begin{eqnarray}
\langle \Psi^{\dag}(\textbf{p}_{1}) \Psi (\textbf{p}_{1})\rangle_{N}
=
\sum_{s=1}^{N}\frac{Z_{N-s}^{0}}{Z_{N}^{0}}\Phi_{1}(\textbf{k},\textbf{q},\beta
\omega s),
 \label{60} \\
 \langle \Psi^{\dag}(\textbf{p}_{2}) \Psi (\textbf{p}_{2})\rangle_{N}
= \sum_{s=1}^{N}\frac{Z_{N-s}^{0}}{Z_{N}^{0}}\Phi_{1}(\textbf{k}, -
\textbf{q},\beta \omega s) ,  \label{60.1}
\end{eqnarray}
where we introduced shorthand notation
\begin{eqnarray}
\Phi_{1}(\textbf{k},\textbf{q},\beta \omega s) =
\frac{b^{3}}{(2\pi\sinh(\beta \omega s))^{3/2}}\exp\left ( -
\left(\textbf{k}-
\frac{1}{2}\textbf{q}\right)^{2}b^{2}\tanh(\frac{1}{2}\beta \omega
s) \right ).
 \label{61}
\end{eqnarray}
Utilizing   Eq. (\ref{47}) and the same technique which was  used to
derive $\langle \Psi^{\dag}(\textbf{r}_{1}) \Psi
(\textbf{r}_{2})\rangle_{N}$,  we get after
somewhat lengthy but straightforward calculations,
\begin{eqnarray}
\langle
\Psi^{\dag}(\textbf{p}_{1})\Psi^{\dag}(\textbf{p}_{2})\Psi(\textbf{p}_{1})\Psi(\textbf{p}_{2})\rangle_{N}=\nonumber
\\
\sum_{s=1}^{N-1}\sum_{s'=1}^{N-s}\frac{Z_{N-s-s'}^{0}}{Z_{N}^{0}}\left (
\Phi_{1}(\textbf{k},  \textbf{q},\beta \omega s)\Phi_{1}(\textbf{k},
- \textbf{q},\beta \omega s') + \Phi_{2}(\textbf{k},
\textbf{q},\beta \omega s)\Phi_{2}(\textbf{k}, - \textbf{q},\beta \omega
s')\right ), \label{62}
\end{eqnarray}
where we introduced notation
\begin{eqnarray}
\Phi_{2}(\textbf{k},\textbf{q},\beta \omega s) =
\frac{b^{3}}{(2\pi\sinh(\beta \omega s))^{3/2}}\exp\left ( -
\textbf{k}^{2}b^{2}\tanh(\frac{1}{2}\beta \omega s ) - \textbf{q}^{2}\frac{b^{2}}{4\tanh(\frac{1}{2}\beta
\omega s)} \right ).
 \label{63}
\end{eqnarray}
Inserting  Eqs. (\ref{60}),  (\ref{60.1}) and (\ref{62}) in Eq.
(\ref{59}) gives us an explicit expression for the  two-boson momentum
correlation function at fixed multiplicities, 
\begin{eqnarray}
C_{N}(\textbf{k},\textbf{q})=
G_{N} \frac{\sum_{s=1}^{N-1}\sum_{s'=1}^{N-s}\frac{Z_{N-s-s'}^{0}}{Z_{N}^{0}}
\Phi_{1}(\textbf{k},  \textbf{q},\beta \omega s)\Phi_{1}(\textbf{k},
- \textbf{q},\beta \omega s')}{\sum_{s=1}^{N}\frac{Z_{N-s}^{0}}{Z_{N}^{0}}\Phi_{1}(\textbf{k},\textbf{q},\beta
\omega s)\sum_{s'=1}^{N}\frac{Z_{N-s'}^{0}}{Z_{N}^{0}}\Phi_{1}(\textbf{k},-\textbf{q},\beta
\omega s')} +\nonumber
\\ G_{N} \frac{\sum_{s=1}^{N-1}\sum_{s'=1}^{N-s}\frac{Z_{N-s-s'}^{0}}{Z_{N}^{0}}
 \Phi_{2}(\textbf{k},
\textbf{q},\beta \omega s)\Phi_{2}(\textbf{k}, - \textbf{q},\beta \omega
s')}{\sum_{s=1}^{N}\frac{Z_{N-s}^{0}}{Z_{N}^{0}}\Phi_{1}(\textbf{k},\textbf{q},\beta
\omega s)\sum_{s'=1}^{N}\frac{Z_{N-s'}^{0}}{Z_{N}^{0}}\Phi_{1}(\textbf{k},-\textbf{q},\beta
\omega s')}.
\label{63.1}
\end{eqnarray}
To estimate normalization
constant $G_{N}$ in Eq. (\ref{63.1}), one needs to utilize the limit $|\textbf{q}| \rightarrow
\infty$ at fixed $\textbf{k}$ in the corresponding expression. One
can readily see that when $|\textbf{q}| \rightarrow \infty$ at fixed
$\textbf{k}$ then 
$C_{N}(\textbf{k},\textbf{q}) \rightarrow
G_{N}\frac{Z_{N-2}^{0}}{Z_{N}^{0}}\left (\frac{Z^{0}_{N}}{Z^{0}_{N-1}}\right )^{2}$. It follows
then that  proper normalization is reached if 
\begin{eqnarray}
G_{N}=
\frac{Z_{N}^{0}}{Z_{N-2}^{0}}\left(\frac{Z^{0}_{N-1}}{Z^{0}_{N}}\right
)^{2}.
 \label{64}
\end{eqnarray}

\section{Results and discussion}

In this section, we  calculate  one-particle momentum spectra and two-particle Bose-Einstein
momentum correlations in the model.  For specificity, we assume  that $m$ is equal to pion mass
and   we take the  set of parameters corresponding
roughly to the values at the system's breakup in $p+p$ collisions   at the LHC
energies: The  temperature $T$ is set to $150$ MeV,  and for  $R$  
we use   $1.5$  and $3$ fm. The thermal wavelength $\Lambda_{T}= 1/\sqrt{mT}\approx 1.36$ fm.  We varied $N$ in the range $1,...,20$. 
Our aim here is to investigate  how particle momentum spectra and correlations in the 
canonical ensemble with the fixed particle number constraint differ from the ones in the corresponding 
grand-canonical ensemble. 

\begin{figure}[!ht]
\centering
\includegraphics[scale=0.5]{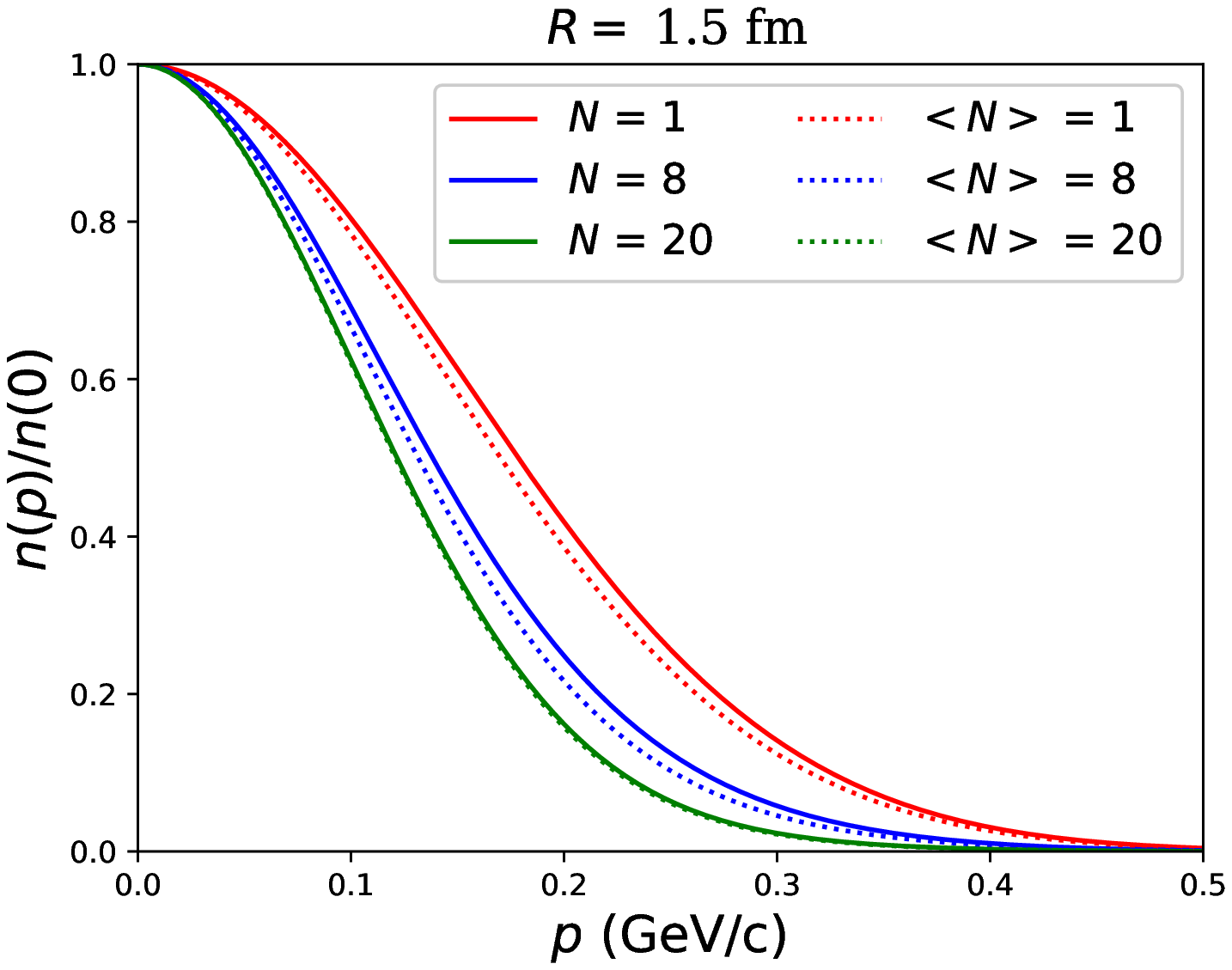}
 \includegraphics[scale=0.5]{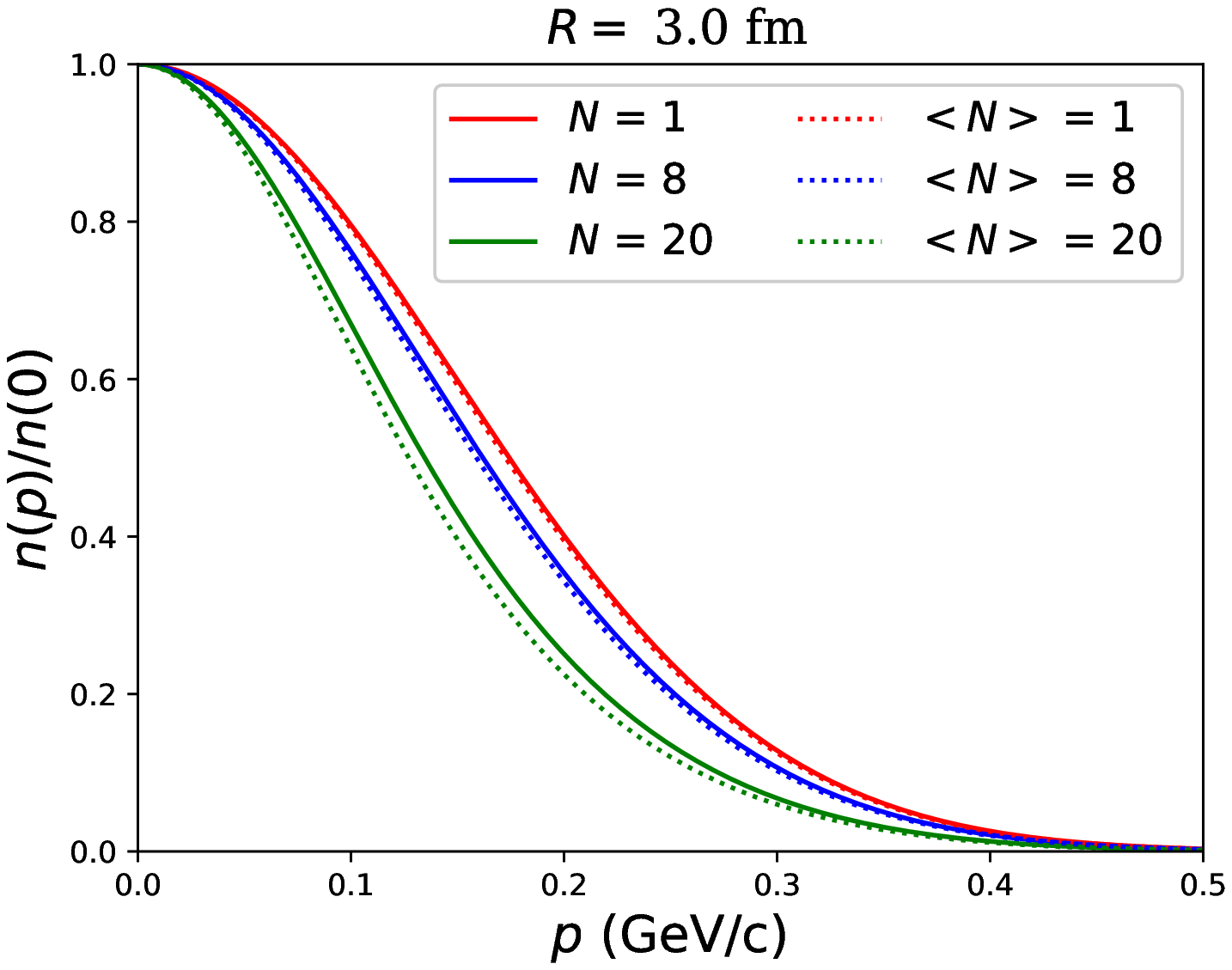}
\caption{Normalized $n(p_{x},0,0)/n(\textbf{0})$   momentum spectra   calculated  in the canonical ensembles with different  $N$  and  $R$ (solid lines), and corresponding spectra calculated in  the grand-canonical ensembles  with $\langle N \rangle =N$ (dotted lines).}
\label{fig:spectra}
\end{figure}

We start with calculations of the one-particle momentum spectra in the canonical ensemble,  
$n_{N}(\textbf{p}) \equiv \langle \Psi^{\dag}(\textbf{p}) \Psi (\textbf{p})\rangle_{N}$;  see Eq. (\ref{60}). 
We compare these calculations
with the ones performed in the corresponding grand-canonical ensembles where $\hat{\mu}$ 
were found numerically to guarantee proper values of $\langle N\rangle$,  such as   $\langle N\rangle = N$. 
One-particle momentum spectra in the grand-canonical ensembles 
are calculated utilizing Eq. (\ref{60}) after substitution $\sum_{s=1}^{N}\frac{Z_{N-s}^{0}}{Z_{N}^{0}} \rightarrow
\sum_{s=1}^{\infty}e^{\beta \hat{\mu} s}$.  The
results are plotted in Fig. \ref{fig:spectra} as a function of the particle momentum 
for several different values of the 
radius parameter $R$ and  particle number $N$. Figure \ref{fig:spectra} demonstrates clearly that
for the used range of parameter values,  
 one-particle momentum spectra in the canonical ensembles can be approximated with good accuracy 
 by the ones calculated in the 
 corresponding grand-canonical ensembles. 
 
\begin{figure}[!ht]
\centering
\includegraphics[scale=0.5]{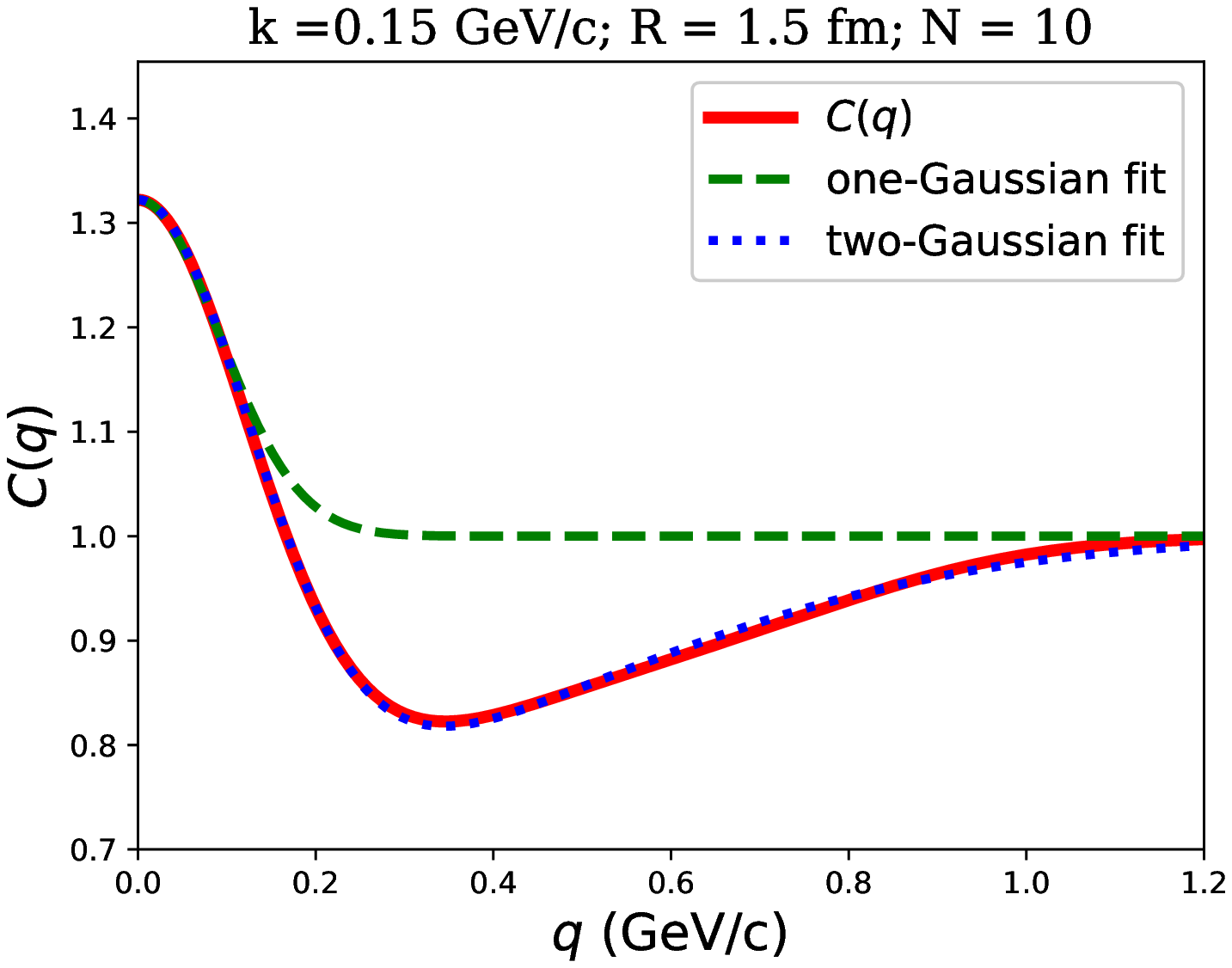}
 \includegraphics[scale=0.5]{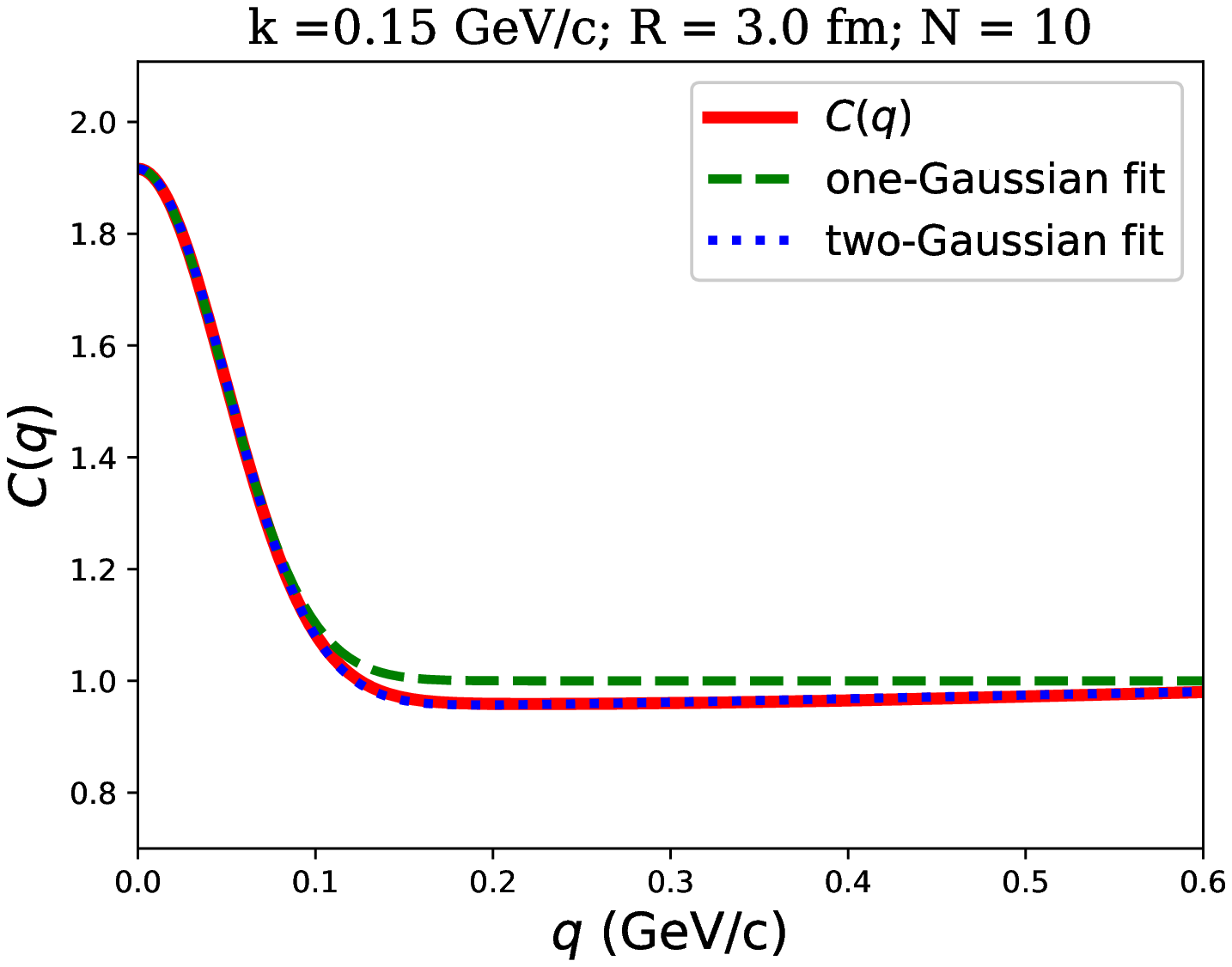}
\caption{Correlation functions (red solid lines) and their one- and two-Gaussian fits (blue dotted  and green dashed lines, respectively) with  $k=0.15$ GeV/c, $N=10$,   $R=1.5$ fm (left plot) and $R=3.0$ fm (right plot). See  text for details.} 
\label{fig:CF}
\end{figure}

Figure \ref{fig:CF} displays two-boson momentum correlation functions (\ref{63.1}) calculated in the canonical ensembles
 as a function of the momentum difference.  From Fig. \ref{fig:CF} it is evident  that 
 the intercept of the correlation function, $C_{N}(\textbf{k},\textbf{0})$, is less than $2$. 
 This can be interpreted as a result of partial coherence of  particle emission \cite{Sin-1} because   projection of the
thermal grand-canonical ensemble into the fixed-$N$ subensemble results in the 
$N$-particle canonical state which is the state with partial coherence. 
 Furthermore, one observes for small values of $R$  the  essential 
 non-Gaussianity of the  correlation functions  beyond the region
of the correlation peak. It distinguishes 
 two-boson correlation functions in the canonical ensembles from the ones in the corresponding 
 grand-canonical ensembles  where
 the correlation functions (not shown here) are well fitted by the Gaussian and intercept of the ones  is equal to $2$.

To analyze reasons for this  behavior of the correlation functions  in greater detail, let us first remark that 
correlation function $C_{N}(\textbf{k},\textbf{q})$ [see Eqs. (\ref{63.1})  and  (\ref{64})]  can be parametrized by the two-Gaussian  expression
\begin{eqnarray}
C_{N}^{2g}(\textbf{k},\textbf{q})= 1- \lambda_{1}(\textbf{k},N)e^{-\textbf{q}^{2}R_{1}^{2}(\textbf{k},N)}+\lambda_{2}(\textbf{k},N)e^{-\textbf{q}^{2}R_{2}^{2}(\textbf{k},N)},
\label{65}
\end{eqnarray}
where $\lambda_{1}>0$ and   $\lambda_{2}>0$.
Here $1- \lambda_{1}(\textbf{k},N)e^{-\textbf{q}^{2}R_{1}^{2}(\textbf{k},N)}$ is  associated with
the first term in Eq. (\ref{63.1}), and $\lambda_{2}(\textbf{k},N)e^{-\textbf{q}^{2}R_{2}^{2}(\textbf{k},N)}$ 
with the second one. The results of fittings are plotted in Fig. \ref{fig:CF}. It is evident that   $C_{N}(\textbf{k},\textbf{q})$ is rather well fitted by Eq. (\ref{65}).  This  suggests that
much of the non-Gaussian deviations observed in Fig. \ref{fig:CF} arises from such a two-scale structure of the correlation function.  If the fitting procedure is restricted to the correlation  peak region,
then one observes from Fig. \ref{fig:CF} that the correlation function is well fitted by the one-Gaussian expression 
\begin{eqnarray}
C_{N}^{1g}(\textbf{k},\textbf{q})= 1+\lambda(\textbf{k},N)e^{-\textbf{q}^{2}R_{HBT}^{2}(\textbf{k},N)},
\label{66}
\end{eqnarray}
where  $\lambda$ is equal to the intercept of the correlation function,  $C_{N}(\textbf{k},\textbf{0})$.

\begin{figure}[!ht]
\centering
\includegraphics[scale=0.7]{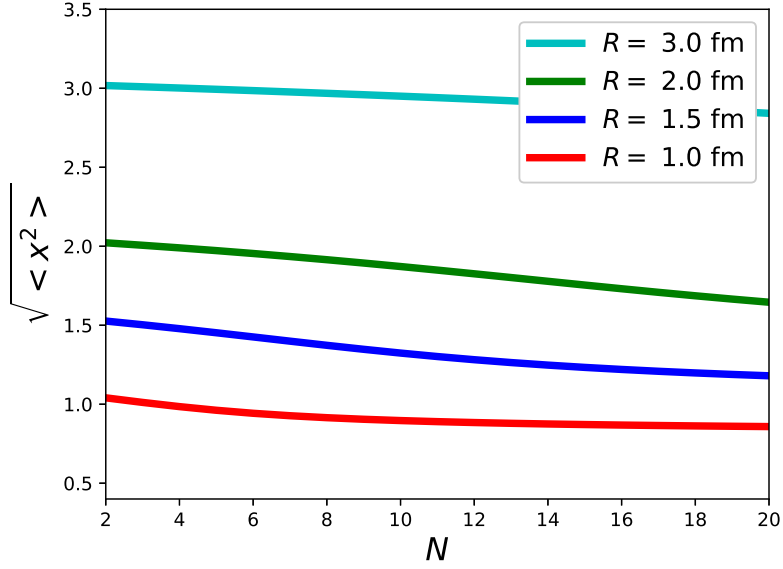}
\caption{The $\sqrt{\langle x^{2} \rangle_{N}}$ dependence on $N$ at different  $R$. }
\label{fig:x2}
\end{figure}

From Fig. \ref{fig:CF} it is clear that the  value of the intercept of the correlation function is 
strongly dependent on the value of $R$ at fixed $N$, namely, one observes   that smaller values of $R$ result in 
 smaller values of the intercept of the correlation function.
The question naturally arises: why does decreasing the parameter $R$ amount  to a decreasing of the intercept? 
Some insight into this question may be gained from
Fig. \ref{fig:x2}, in which mean size of the system $\sqrt{\langle x^{2} \rangle_{N}}$ [see Eq. (\ref{55})]
is plotted out to $N$. One observes from this figure  that parameter $R$ roughly corresponds to the mean spatial 
size of the system in the varied range of $N$. It means that the  decrease of $R$ at fixed $N$ results 
in an increase of the mean particle number density, $\varpropto  N/R^{3}$. 

\begin{figure}[!ht]
\centering
\includegraphics[scale=0.7]{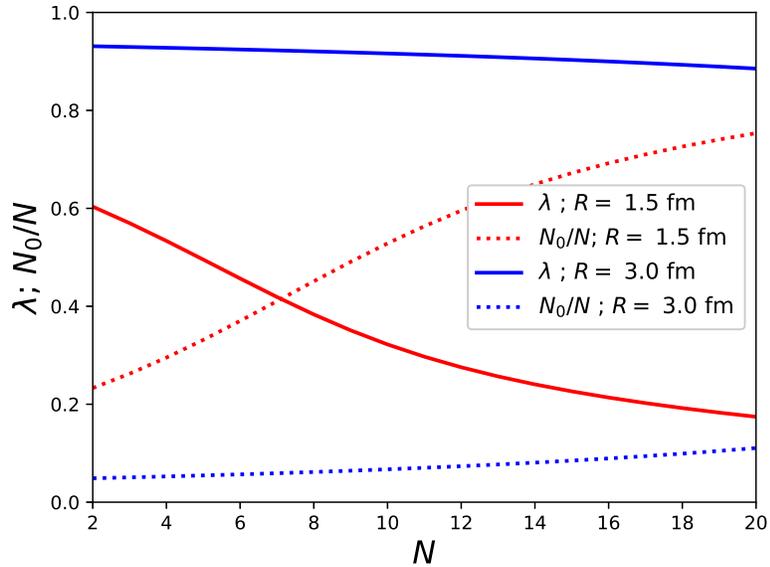}
\caption{The $\lambda$ at $ k=0.15 $ GeV/c,   and $N_{0}/N$ dependence on  $N$ for  $R=1.5$ fm and $R=3.0$ fm.  }
\label{fig:lambda}
\end{figure}

\begin{figure}[!ht]
\centering
\includegraphics[scale=0.5]{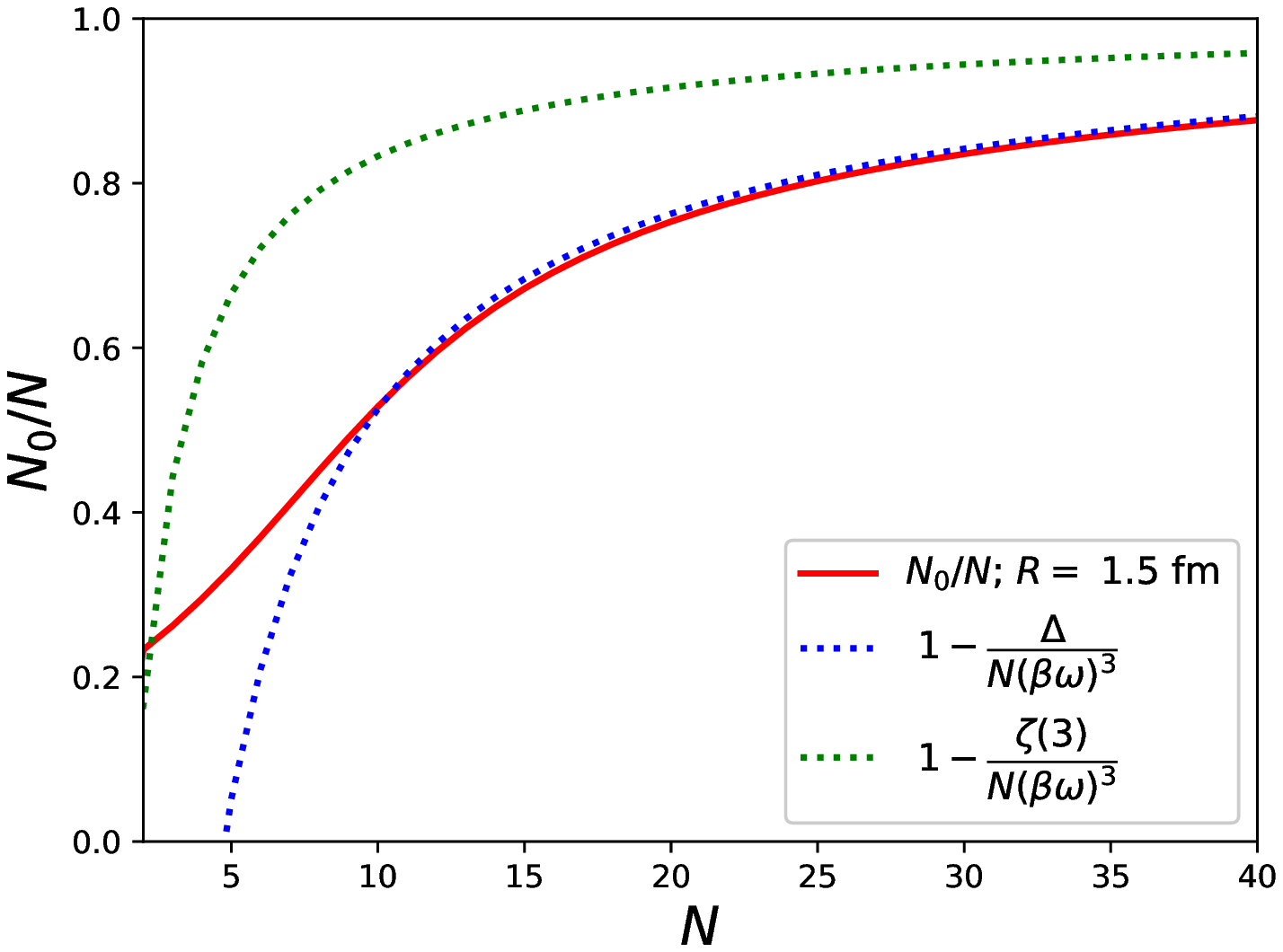} 
\includegraphics[scale=0.5]{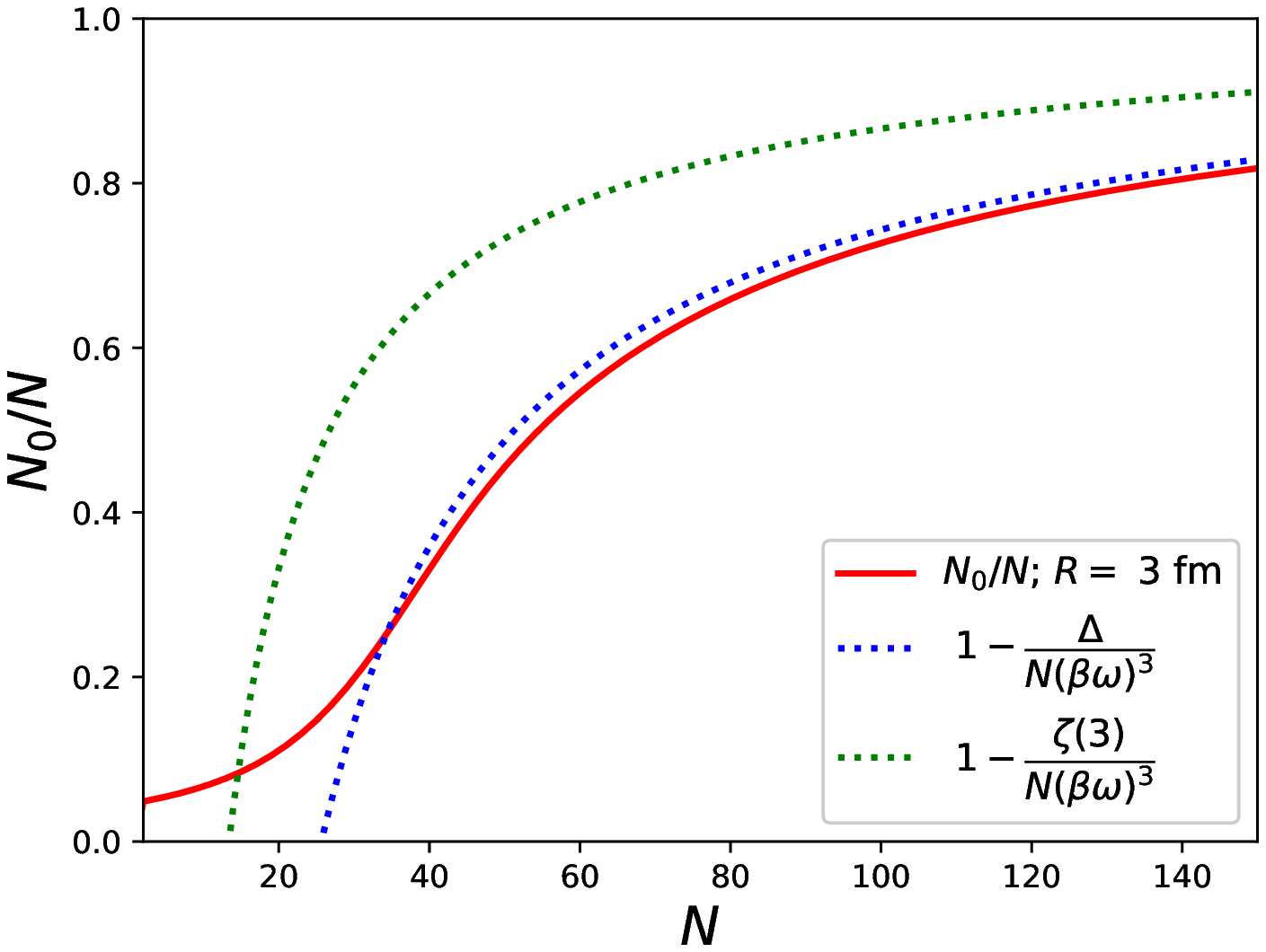}
\caption{Canonical $N_{0}/N$ (red solid line) and its fits with Eq. (\ref{69}) (blue and green dotted lines), $R=1.5$ fm (left plot), and $R=3.0$ fm (right plot). See  text for details.}
\label{N1}
\end{figure}

To gain further insight into these results, the $\lambda$ parameter and also the ratio
of the ground-state
population, $N_{0}=\langle
\alpha^{\dag}(\textbf{0})\alpha(\textbf{0})\rangle_{N}$ [see Eq. (\ref{43})], 
to the number of particles, $N$,  are plotted out to $N$ in Fig. \ref{fig:lambda}. 
It can be seen from this figure that the coherent effects, associated with the
parameter $\lambda$, are significant for any $N$ if the mean size of the system  
is comparable to or less than  the thermal wavelength $\Lambda_{T}$. 
One can also see from this figure that an increase of  $N$   results in 
 an increase of the value of the  $N_{0}/N$ ratio and decrease of the value of
 the $\lambda$ parameter.  To interpret this result it is instructive to compare 
the canonical  condensate fraction, 
 $N_{0}/N$, with its  grand-canonical counterpart
 $\langle N_{0} \rangle / \langle N \rangle$. We start by noting that 
 applying Cauchy's integral formula  to Eqs. (\ref{39.0}) and  (\ref{39.00}) one can get
 (see, e.g., Ref. \cite{formula})
 \begin{eqnarray}
Z_{N}^{0}= \beta \int_{\delta - i \infty}^{\delta + i \infty}\frac{d\hat{\mu}}{2\pi i}  e^{-\hat{\mu} \beta N}  Z (\hat{\mu}).
\label{67}
\end{eqnarray}
 It is well known that utilizing the above expression for approximate evaluation of the canonical partition function, in the leading order of the saddle-point approximation one obtains 
 \begin{eqnarray}
Z_{N}^{0} \approx e^{-\hat{\mu}_{\sigma} \beta N}Z(\hat{\mu}_{\sigma}),
\label{68}
\end{eqnarray}
where $\hat{\mu}_{\sigma}$ is solution of the equation $\frac{d}{d\hat{\mu}}(-\hat{\mu} \beta N + \ln{Z(\hat{\mu})})=0$.
For an ideal gas it means that $\hat{\mu}_{\sigma}$ is such that $\langle N \rangle =N$.   Equation  (\ref{68}) becomes exact for $N \rightarrow \infty$. Then, using Eqs. (\ref{39.0}), (\ref{39.00}),  and  (\ref{40.1}),  one can expect that for finite but large $N$ we get $N_{0}/N \approx \langle N_{0} \rangle /\langle N \rangle  $ where $\langle N_{0} \rangle /\langle N \rangle $
is the condensate fraction in the grand-canonical ensemble with $\langle N \rangle =N$.
Let us compare our  results for the canonical  condensate fraction, 
 $N_{0}/N$, with the grand-canonical condensate fraction for a finite mean number of particles in a three-dimensional harmonic potential  \cite{BEC-1},
\begin{eqnarray}
\frac{\langle N_{0} \rangle}{\langle N \rangle} \approx 1-\frac{\Delta}{\langle N \rangle (\beta \omega)^{3}},
\label{69} \\
\Delta = \zeta (3) + \frac{3}{2} \zeta (2) \beta \omega , \label{70}
\end{eqnarray}
calculated  in the approximation $\beta \omega \ll 1$. 
Here $\zeta (x)$ is the Riemann zeta function,  $\zeta (2) \approx 1.645$ and  $\zeta (3) \approx 1.202$. In the above expression we have approximated $e^{\beta(\hat{\mu}-(3/2)\omega)}\approx 1$.  Identifying  $\langle N \rangle $ with the actual particle number $N$, 
and $\beta \omega$ with $\Lambda_{T}/R$, see Eq. (\ref{57}), we compare $N_{0}/N$ with $\langle N_{0} \rangle /\langle N \rangle $ in Fig. \ref{N1}  for $R=1.5$ fm and $R=3$ fm. 

 One observes that the approximate grand-canonical formula  shows a rather good agreement 
 with the exact
canonical results even for not very large values of $N$.
 Loosely speaking, 
 the canonical condensate fraction  of the large system  becomes noticeable (say, about $1/2$)
when the mean interparticle distance, $(N/R^{3})^{-1/3}$, 
becomes smaller than the correlation length, for an ideal gas the latter coincides with the thermal wavelength, $\Lambda_{T}$. 

While the quantitatively
accurate description of the canonical  condensate fraction within the grand-canonical 
approximation is 
manifest, it is not the case for fluctuations.  It is well known  that fluctuations in 
the ground state differ in the canonical and grand-canonical ensembles,
and that for the latter  the  condensate fluctuations  are very large;
see, e.g., Ref. \cite{cond} and references therein.  In the canonical ensemble with 
a fixed  number of
particles such large fluctuations are impossible,  and  therefore an increase
of the ground-state fraction $N_{0}/N$  increases ``coherence''  of the state. 
The latter distinguishes the ideal gas   Bose-Einstein condensation in the canonical  ensemble from 
 the ideal gas Bose-Einstein condensation in the grand-canonical  ensemble. 
 It is well known that  the intercept of the two-boson momentum correlation function
 for a  maximally mixed (chaotic) state is equal to
$2$,  and  that the one  for a pure state is equal to $1$; see,
e.g., Ref. \cite{Sin-1}. Therefore, an  increase of the ground-state fraction, $N_{0}/N$, 
 results in a decreasing of the $\lambda$ parameter.
 
\begin{figure}[!ht]
\centering 
\includegraphics[scale=0.5]{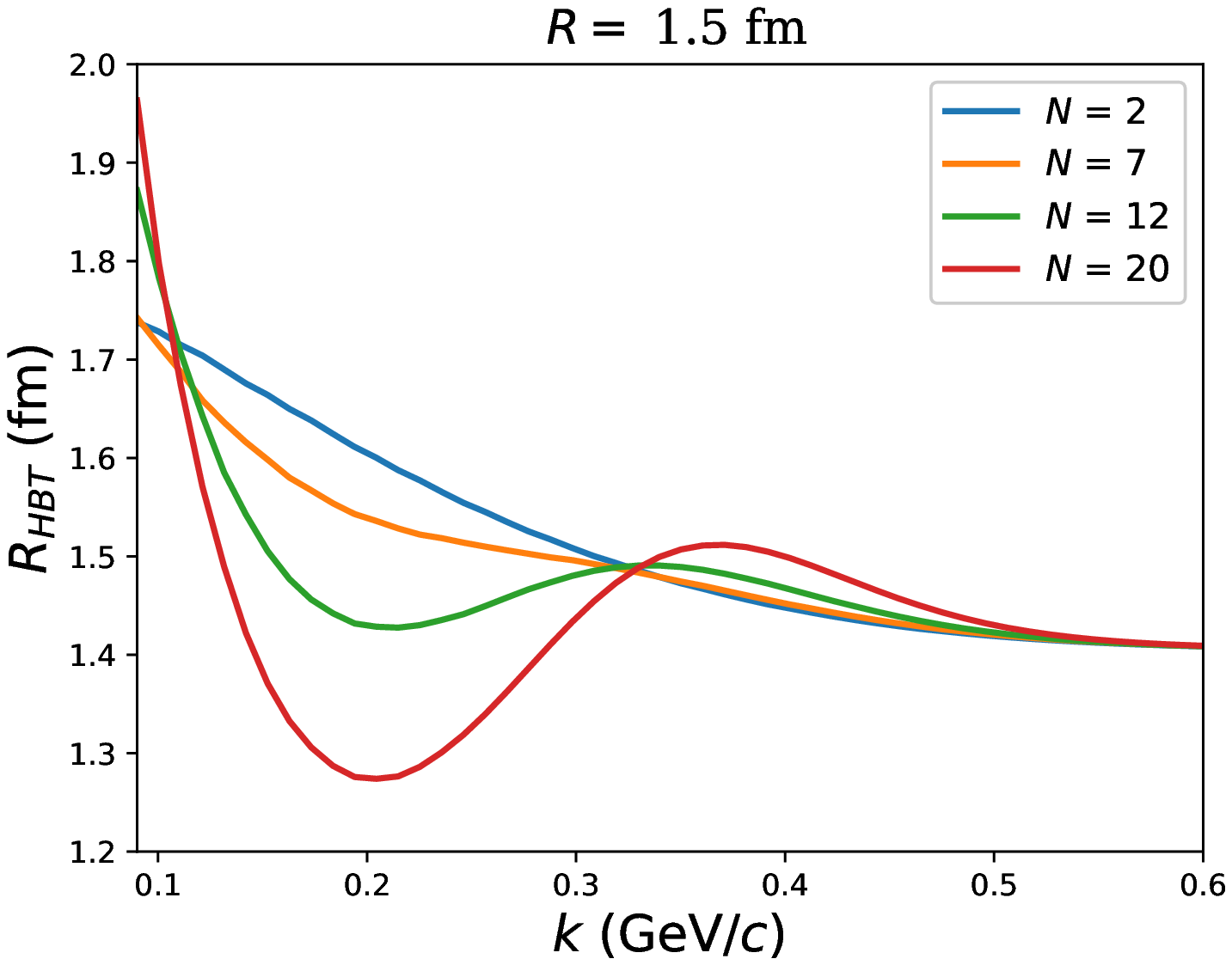} 
\includegraphics[scale=0.5]{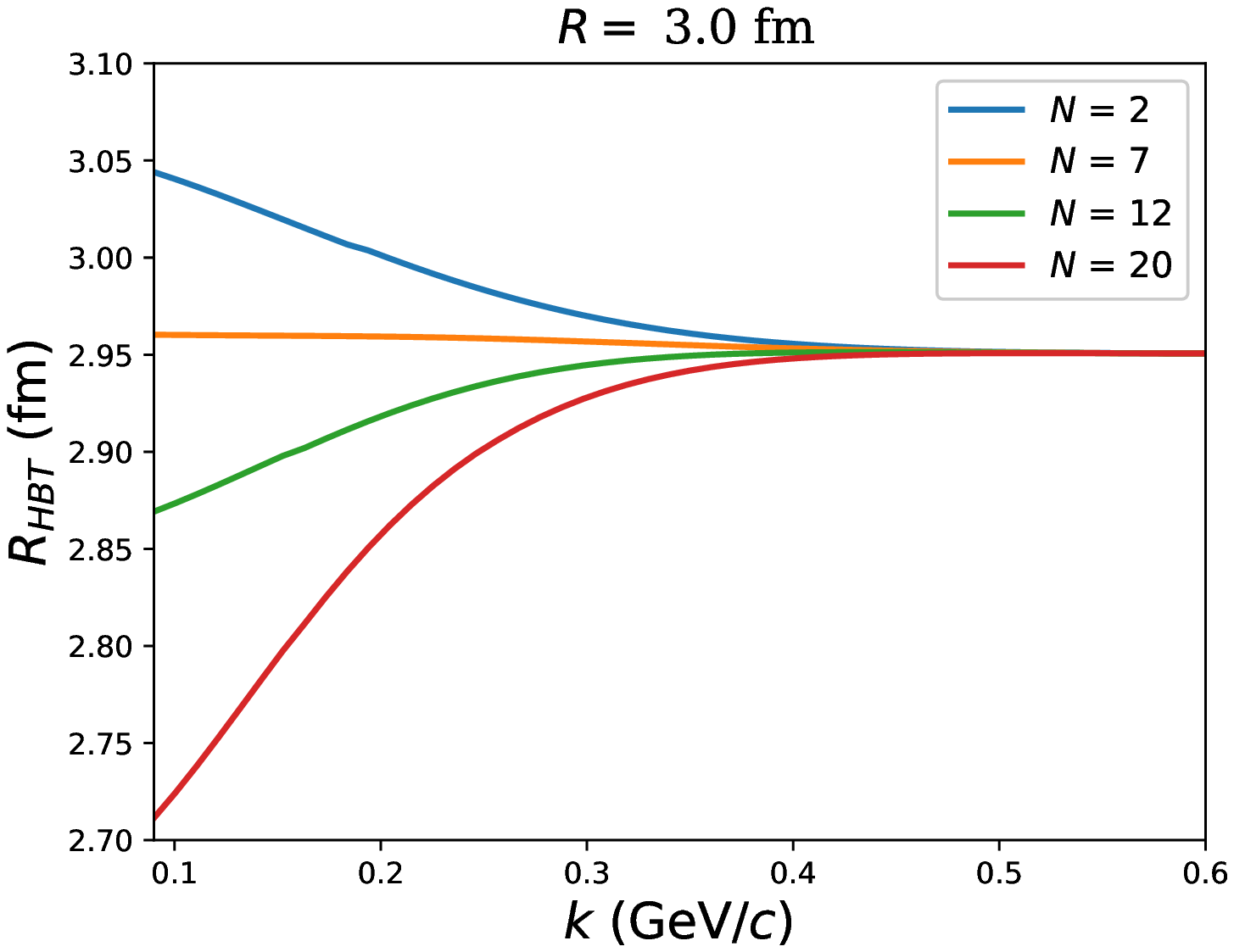}
\caption{HBT radii obtained from the one-Gaussian fit of the two-boson correlation function in the canonical ensembles with different $N$, as a function of the pair average momentum $k$.}
\label{fig:hbt}
\end{figure}

Finally, in Fig. \ref{fig:hbt} we plot the $R_{HBT}$ as a function of the  pair momenta, $k$, for 
different $R$ and $N$. One observes a consistent trend: by increasing  $k$ the interferometry radii,
$R_{HBT}$, become independent of $N$. 

\section{Conclusions}

Usually one does not care so much about quantum coherence in the
canonical ensemble at fixed multiplicities.\footnote{See, however,
Ref. \cite{Sin-3} where it was demonstrated that the description in
the hydrodynamic approach of the interferometry radii in $p+p$
collisions is improved if one accounts for the  mutual quantum coherence of
closely located emitters caused by the uncertainty principle.}
However, utilizing the simple  analytically solvable model, 
we demonstrated   that
the formulas derived in  the  fixed-$N$ canonical ensemble for a small  
inhomogeneous thermal system are not  always    accurately  approximated by 
the grand-canonical ones with $\langle N \rangle =N$. Namely, we noticed that  while the one-particle momentum spectra 
can be well approximated by the corresponding grand-canonical ensemble expressions, it is not the case for the 
 two-boson momentum correlations.
 Interestingly, we observed that the  most significant  deviations arise  if the  particle number 
 density in the canonical ensemble can increase with $N$. In  the considered 
 simple model  it implies that interferometry radii are independent on $N$ at moderately high pair momenta. Then
 for fairly high 
 $N$  the  particle number density  exceeds  some limit
 value  leading to the noticeable  Bose-Einstein
condensation in the corresponding  ground state of the fixed-$N$ canonical ensemble state. 
Such a condensation  strengthens  the  coherence properties of the canonical ensemble state, and results
 in the decreasing of the  intercept of the two-boson momentum correlation function when $N$ increases. 
  This may explain the observed phenomenon of partial quantum
coherence in high-multiplicity   $p+p$ collisions events 
in  fixed multiplicity bins at the LHC energies  \cite{Atlas,CMS}. It would be very interesting
to revisit the results of experimental studies in view of our findings.

The main lesson from this study is that the canonical and 
 grand-canonical ensembles can yield different results for two-boson momentum correlations 
 of particles emitted by small 
 inhomogeneous systems. 
The results of our analysis can be useful to  elucidate  the
influence on the shape of the measured correlation function of both factors: 
an experimental selection of events with fixed multiplicity and the effects of
thermalization and  flow. 
Therefore, determination of the extent to which our results can be generalized
for a  realistic model  of heavy ion and particle collisions could be of great interest.

\begin{acknowledgments}
The research was partially (M.A. and Yu.S.) carried out within NAS of Ukraine priority project ``Fundamental properties of the matter in the relativistic collisions of nuclei and in the early Universe'' (No. 0120U100935). 
\end{acknowledgments}

\end{document}